\documentclass[british,a4paper]{article}
 \pdfoutput=1
\usepackage{babel,amsmath,amssymb,amsthm}
\usepackage{bm}
\usepackage[sans]{dsfont}
\usepackage[mathscr]{euscript}
\usepackage{times}
\usepackage{cite}
\usepackage{mathtools}
\usepackage{hyperref}
\usepackage{a4wide}

\theoremstyle{plain}
\newtheorem{theorem}{Theorem}
\newtheorem{proposition}[theorem]{Proposition}

\theoremstyle{remark}
\newtheorem{remark}{Remark}

\theoremstyle{definition}
\newtheorem{definition}{Definition}
\theoremstyle{definition}

\newcommand{\Tr}{\operatorname{Tr}}
\newcommand{\rmd}{\mathrm{d}}
\newcommand{\rme}{\mathrm{e}}
\newcommand{\rmi}{\mathrm{i}}
\newcommand{\RE}{\mathrm{\,Re\,}}
\newcommand{\IM}{\mathrm{\,Im\,}}

\newcommand{\Rbb}{\mathbb{R}}
\newcommand{\Cbb}{\mathbb{C}}

\newcommand{\T}{\mathtt{T}}

\newcommand{\openone}{\mathds{1}}

\newcommand{\ind}{\mathtt{1}}
\newcommand{\id}{\mathrm{Id}}

\newcommand{\abs}[1]{\left\vert#1\right\vert}

\newcommand{\Ical}{\mathcal{I}}

\newcommand{\Kcal}{\mathcal{K}}
\newcommand{\Lcal}{\mathcal{L}}

 \newcommand{\Rcal}{\mathcal{R}}

\newcommand{\Tcal}{\mathcal{T}}

 \newcommand{\Wcal}{\mathcal{W}}

\newcommand{\Bscr}{\mathscr{B}}

\newcommand{\Hscr}{\mathscr{H}}

 \newcommand{\Nscr}{\mathscr{N}}

\newcommand{\Tscr}{\mathscr{T}}

\begin{document}

\title{Markovian master equations for quantum-classical hybrid systems}

\author{ALBERTO BARCHIELLI
\\
Istituto Nazionale di Fisica Nucleare (INFN), Sezione di Milano,Italy; \\ also
Istituto Nazionale di Alta Matematica (INDAM-GNAMPA)
\\ alberto.barchielli@polimi.it
}

\maketitle

\begin{abstract}
The problem of constructing a consistent quantum-classical hybrid dynamics is afforded in the case of a quantum component in a separable Hilbert space and a continuous, finite-dimensional classical component. In the Markovian case, the problem is formalized by the notion of \emph{hybrid dynamical semigroup}. A classical component can be observed without perturbing the system and information on the quantum component can be extracted, thanks to the quantum-classical interaction. This point is formalized by showing how to introduce \emph{positive operator valued measures} and \emph{operations} compatible with the hybrid dynamical semigroup; in this way the notion of hybrid dynamics is connected to quantum measurements in continuous time. Then, the case of the most general \emph{quasi-free} generator is presented and the various quantum-classical interaction terms are discussed. To bee quasi-free means to send, in the Heisenberg description, hybrid Weyl operators into multiples of Weyl operators; the results on the structure of quasi-free semigroups were proved in Reference \cite{BarW23}. Even in the pure quantum case, a quasi-free semigroup is not restricted to have only a Gaussian structure, but also jump-type terms are allowed. An important result is that,  to have interactions producing a flow of information from the quantum component to the classical one, suitable dissipative terms must be present in the generator. Finally, some possibilities are discussed to go beyond the quasi-free case.
\end{abstract}



\section{Introduction}\label{sec:intro}
The search for a consistent formulation of the dynamics for quantum-classical hybrid systems has a long history; the motivations include
computational advantages, description of mesoscopic systems, unification of gravity and quantum theory, description of quantum measurements\ldots, see \cite{DGS00,Dio14,OppWD22,Sergi+23,Dio23,ManRT23,Pomar+23,Opp+23,ProsB23,DamWer23,BarL05,BarW23} and references there in. Also the quantum measurements in continuous time can be interpreted in terms of hybrid systems; in this case the classical component
is the monitored signal extracted from the quantum system \cite{Dio14,Dio23,BarB91,BarHL93,BarP96,BarG09}. A typical realization is given by direct, homodyne and heterodyne detection in quantum optics  \cite{BarG13,Bar06}.

A first aim of this paper is to present the results of \cite{BarW23}, where the most general hybrid Markovian dynamics has been obtained in the quasi-free case. The notion of quasi-free map means that, in the Heisenberg picture, quantum-classical Weyl operators are mapped to multiples of Weyl operators \cite{Van78,DVV79,Hell10,DamWer23}. A quasi-free dynamics includes not only the Gaussian case, but also contributions of ``jump type''.
A second aim is to discuss the physical meaning of the various terms appearing in the master equation giving the dynamics of the hybrid system. An important point here is that, to have flow of information from the quantum system to the classical one, suitable dissipation terms must be present in the dynamics. As the classical component can be observed without disturbing it, this means that at least some ``smooth'' state reduction must be present to have significant quantum-classical interactions, allowing extraction of quantum information.

The general setting of the hybrid dynamics is discussed in Sec.\ \ref{sec:hds}. The quantum component is described in a separable Hilbert space and the classical one is continuous and finite dimensional; the dynamics of the total system is without memory. This is formalized by the definition of \emph{hybrid dynamical semigroup}; we work in the Heisenberg description, which is more convenient for the study of the generator of this semigroup. Then, it is shown that the possibility of observing the classical subsystem, even in continuous time, allows to construct \emph{positive operator valued measures} and \emph{instruments}, typical notions in quantum measurement theory.

In Sec.\ \ref{sec:qfsemig} we ask the maps in the semigroup to be quasi-free. Here, we define the hybrid Weyl operators and we present the results of \cite{BarW23} about the structure of the most general quasi-free hybrid dynamical semigroup. Then, the generator of the dynamical semigroup is explicitly given; we put in evidence that it can be decomposed in a ``diffusive-like'' component and in a ``jump-like'' one. We show also how the quantum-classical interaction is made up by different structures.

The knowledge of the generator of the semigroup is equivalent to give the master equation of the hybrid system. The meaning of the various terms appearing in the quasi-free dynamics is discussed in Sec.\ \ref{sec:qfdyn}. The resulting dynamical equation connects together a quantum master equation of Lindblad type with a differential equation of Kolmogorov-Fokker-Planck type. A relevant point is the connection between dissipation and flow of information from the quantum system to the classical one. When we ask the vanishing of the dissipative terms either  in the reduced quantum dynamics or in the reduced classical dynamics, the positivity condition implies the vanishing of the interaction terms giving this flow of information.

Comparisons with other approaches and suggestions for possible developments are given in Sec.\ \ref{sec:compare}.

\section{Quantum-classical dynamical semigroup}\label{sec:hds}

We assume the quantum component to be represented in a separable complex Hilbert space $\Hscr$. Moreover, the bounded linear operators are denoted  by  $\Bscr(\Hscr)$ and the trace class by $\Tscr(\Hscr)$. The unit element in $\Bscr(\Hscr)$ is denoted by $\openone$ and the adjoint of the operator $a\in\Bscr(\Hscr)$ is $a^\dagger$. The complex conjugated of $\alpha\in \Cbb$ is denoted by $\overline \alpha$.

The classical component is a continuous system with \emph{phase space} $\Xi_0:=\Rbb^s$. The probability densities are in $L^1(\Rbb^s)$, whose dual space is $L^\infty(\Rbb^s)$; the Lebesgue measure is always understood.

The observables of the composed system live in the $W^*$-algebra $\Nscr=\Bscr(\Hscr)\otimes L^\infty(\Rbb^s)$, and the hybrid states in its predual $\Nscr_*=\Tscr(\Hscr)\otimes L^1(\Rbb^s)$. The duality form between $\Nscr_*$ and $\Nscr$ is given by
\begin{equation}\label{dualityform}
\langle P|F\rangle= \int_{\Rbb^s}\rmd x\,\Tr\left\{P(x)F(x)\right\}, \qquad \forall P\in \Nscr_*, \qquad \forall F\in \Nscr;
\end{equation}
we have used the fact that $\Nscr_*$ is isomorphic to $L^1\big(\Rbb^s;\Tscr(\Hscr)\big)$ (the Lebesgue integrable functions from $\Rbb^s$ to $\Tscr(\Hscr)$), and $\Nscr$ to $L^\infty\big(\Rbb^s;\Bscr(\Hscr)\big)$ (the bounded functions from $\Rbb^s$ to $\Bscr(\Hscr)$).

\begin{remark} \label{rem:state} A state $\hat\pi$ is a trace-class valued function $\hat\pi(x)\in \Tscr(\Hscr)$, $x\in \Rbb^s$, such that  $\int_{\Rbb^s}\rmd x\, \Tr\{\hat\pi(x)\}=1$ and $\hat\pi(x)\geq 0$. It is always possible to decompose a hybrid state as a probability density times a conditional quantum state \cite{Dio14,Dio23}: (almost everywhere with respect to the Lebesgue measure) we can write
\begin{equation}\label{pxrho}
\hat \pi(x)=p_{\hat \pi}(x) \rho_{\hat \pi}(x), \qquad p_{\hat \pi}(x)=:\Tr\{\hat\pi(x)\}, \qquad \rho_{\hat \pi}(x)\geq 0, \qquad \Tr\left\{\rho_{\hat \pi}(x)\right\}=1.
\end{equation}
Then, $p_{\hat \pi}(x)$ is a probability density representing the reduced state of the classical component and $\rho_{\hat \pi}(x)$ is the quantum state conditional on the value $x$ taken by the classical component. Moreover, $\int_{\Rbb^s}\rmd x\, \hat\pi(x)$ is the statistical operator representing the reduced state of the quantum component.
\end{remark}

To study the dynamical semigroups for the hybrid system it is convenient to work with operators on the $W^*$-algebra $\Nscr$, which means to work in the analogous of the Heisenberg picture of the dynamics.

\begin{definition}\label{def:1} A \emph{hybrid dynamical semigroup} is a family $\{\Tcal_t, \; t\geq 0\}$ of linear bounded operators on $\Nscr$, such that, $\forall t,s\geq 0$,
\begin{itemize}
\item[a.] $\Tcal_t$ is completely positive;
\item[b.] $\Tcal_0=\id$;
\item[c.] $\Tcal_t[\openone]=\openone$;
\item[d.] $\Tcal_t$ is a normal map on $\Nscr$;

\item[e.] $\Tcal_t\circ \Tcal_s=\Tcal_{t+s}$;
\item[f.] $\Tcal_t$ is weak$^*$ continuous in $t$, i.e. $
\langle P|\Tcal_t[F]\rangle$ is continuous in $t$, \ $\forall P\in \Nscr_*$, \ $\forall F\in \Nscr$.
\end{itemize}
\end{definition}

Property a.\ (complete positivity) means that, for any choice of the integer $N\geq 1$, one has
\[
\sum_{i,j=1}^N\int_{\Rbb^s}\rmd x\, \langle g_j(x)| \Tcal_t[F_j^*F_i](x)g_i(x)\rangle\geq 0 \qquad \forall F_j\in L^\infty\big(\Rbb^s;\Bscr(\Hscr)\big), \quad g_j\in \Hscr\otimes L^2(\Rbb^s),
\]
$F^*$ is the adjoint of $F$  as element of $\Nscr\equiv L^\infty\big(\Rbb^s;\Bscr(\Hscr)\big)$. In property b., $\id$ denotes the identity map.
Property d.\ is a suitable regularity requirement \cite{BarHL93,BarP96}); for a positive map it is equivalent to require the map $\Tcal_t$ on $\Nscr$ to be the adjoint of a bounded map ${\Tcal_t}_*$ on $\Nscr_*$.

\begin{remark} \label{rem:Tstar} The dynamics of the quantum-classical system is given by the pre-adjoint semigroup: if $\hat \pi_0$ is the hybrid state at time $0$, then the state at time $t$ is given by $\hat \pi_t={\Tcal_t}_*[\hat \pi_0]$.
\end{remark}

\subsection{Instruments and probabilities}\label{sec:instr}
In principle, a classical system can be observed without disturbing it. Indeed, as done in \cite[Sec.\ 5]{BarW23}, in the pure classical case ($\Hscr=\Cbb$) from the single-time probabilities given at all times $t\geq 0$ one can obtain also the multi-time probabilities, the transition probabilities\ldots; due to the semigroup request one obtains a Markov process \cite[Sec.\ 10]{Sato99}. On the other side, measurements on a quantum system \cite{Hol01,WisM10} can be interpreted as involving hybrid systems: a \emph{positive operator valued measure} is a channel from a quantum system to a classical one, an \emph{instrument} is a channel from a quantum system to a hybrid system \cite{BarL05,Dio14,DamWer23,ManRT23}.

\begin{definition} \label{def:instr} An \emph{instrument} $\Ical(\cdot)$ with value space $\Rbb^s$ is a function from the $\sigma$-algebra of the Borel sets in $\Rbb^s$ to the linear maps on $\Bscr(\Hscr)$ such that
\begin{enumerate}
\item (positivity) for every Borel set $E\subset \Rbb^s$, $\Ical(E)$ is a completely positive and normal map from $\Bscr(\Hscr)$ into itself;
\item (normalization) $\Ical(\Rbb^s)[\openone]=\openone$;
\item ($\sigma$-additivity)  we have $\Ical\left(\bigcup_i E_i\right)[a]=\sum_i\Ical(E_i)[a]$, $\forall a\in \Bscr(\Hscr)$ and for every countable family of Borel disjoint sets $E_i$.
\end{enumerate}\end{definition}
Let us recall that an instrument gives the probabilities and the conditional state after the measurement. The quantity $\Ical(\cdot)[\openone]$ is a positive operator valued measure, and it gives only the probabilities.

As in the purely classical case, in the hybrid case too the classical component can be observed in continuous time without disturbance. By analogy with the transition probabilities, it is possible to introduce instruments depending on the initial value of the classical system, some kind of \emph{transition instruments} \cite{BarH95,BarW23}. Indeed,
we can define the family of maps
\begin{equation}\label{I|B}
\Ical_t(E|x)[a]=\Tcal_t[a\otimes \ind_E](x), \qquad \forall a\in\Bscr(\Hscr);
\end{equation}
$E\subset \Rbb^s$ is a generic Borel set. Definition \eqref{I|B} holds almost everywhere for $x\in \Rbb^s$.  By $\ind_E$ we denote the \emph{indicator function} of a generic set $E$:
\[
\ind_E(x)=\begin{cases} 1 &\text{if } \ \ x\in E,
\\
0 &\text{if } \ \ x\notin E.\end{cases}
\]

\begin{proposition}\label{prop:transinstr} Almost everywhere for $x\in \Rbb^s$,
equation \eqref{I|B} defines an instrument $\Ical_t(\cdot|x)$ on the $\sigma$-algebra of the Borel sets in $\Rbb^s$.
Moreover, the family of instruments \eqref{I|B} enjoys  the following composition property:
\begin{equation}\label{qCKeq}
\Ical_{t+t'}(E|x)=\int_{z\in \Rbb^s} \Ical_{t'}(\rmd z|x)\circ\Ical_t(E|z).
\end{equation}
\end{proposition}
The proof of this proposition is given in \cite{BarW23}.  The quantity $\Ical_t(\bullet|x)[\openone]$ turns out to be a positive operator valued measure, conditional on $x$.
Equation \eqref{qCKeq} is the quantum analogue of the Chapman-Kolmogorov identity for transition probabilities \cite[Sec.\ 10]{Sato99}.

Property 4.\ represents a compatibility condition among the various instruments at different times. Via Kolmogorov's extension theorem \cite[Theor.\ 1.8]{Sato99}, this property allows to represent the classical component as a stochastic process $X(t)$, whose joint probabilities at the times $0< t_1<t_2<\cdots<t_m$ (for an initial state $\hat\pi_0$) are given by
\begin{multline}\label{genprobs}
P[X(t_1)\in E_1, X(t_2)\in E_2, \ldots, X(t_m)\in E_m|\hat\pi_0]
=\int_{\Rbb^s}\rmd x \Tr \biggl\{\hat\pi_0(x) \int_{E_1}\Ical_{t_1}(\rmd x_1|x)\\ {}\circ \int_{E_2}\Ical_{t_2-t_1}(\rmd x_2|x_1)\circ \cdots
\circ \int_{E_m}\Ical_{t_m-t_{m-1}}(\rmd  x_m|x_{m-1})[\openone]\biggr\}.
\end{multline}

\section{Quasi-free hybrid dynamics}\label{sec:qfsemig}
When the dynamical semigroup is restricted to quasi-free maps, its structure can be completely characterized \cite{BarW23}.
As said in Sec.\ \ref{sec:intro}, quasi-free maps are defined by their action on the Weyl operators; to introduce such operators also the quantum system is taken to be continuous with Hilbert space
\begin{equation}
\Hscr=L^2(\Rbb^n).
\end{equation}
\subsection{Settings}
Firstly, we introduce the position  and momentum operators $Q_j$, $P_j$; we also use the vector notation
\begin{equation}\label{R=QP}
R=\begin{pmatrix}Q\\ P\end{pmatrix}, \qquad R_j =\begin{cases} Q_j   & j=1,\ldots, n,
\\ P_{j-n} \quad    & j= n+1 ,\ldots, 2n.
\end{cases}
\end{equation}
The canonical commutation relations (CCR) take the form
\begin{equation*}
[R_i,R_j]=\rmi\sigma_{ij}, \qquad \sigma_{ij}=\begin{cases}1  & 1\leq i\leq n\quad j=i+n,
\\ -1 \quad & n+1\leq i \leq 2n \quad j=i-n,
\\ 0 & \text{otherwise},\end{cases}
\end{equation*}
and the Weyl operators can be written as
\begin{subequations}
\begin{equation}
W_{1}(\zeta)=\exp \left\{\rmi \zeta \cdot R\right\}\in \Bscr(\Hscr), \qquad \zeta\in \Xi_{1}:=\Rbb^{2n};
\end{equation}
$\Xi_{1}$ is the \emph{quantum phase space}.

For the classical component, the analogous objects  are the Weyl functions:
\begin{equation}
W_{0}(k)\in L^\infty(\Rbb^s), \qquad W_{0}(k)(x)= \exp\left\{\rmi  k\cdot x\right\},\quad k,\, x\in \Xi_{0}.
\end{equation}
For the hybrid system we can introduce the total phase space $\Xi$ and the (generalized) Weyl operators $W(\xi)$:
\begin{equation}\begin{split}
&\Xi=\Xi_{1} \oplus \Xi_{0}=\Rbb^{d}, \qquad d=2n +s,
\\ &W(\xi)=  W_{1}(\zeta)W_{0}(k)\in \Nscr, \qquad \xi=\begin{pmatrix} \zeta \\ k \end{pmatrix}, \qquad \zeta\in \Xi_1, \qquad k\in \Xi_0.
\end{split}\end{equation}
\end{subequations}
The Weyl operators satisfy the following composition property:
\begin{equation}\label{WxiWxi'}
W(\xi+\eta)
=W(\xi)W(\eta)\exp\left\{\frac\rmi 2\,\xi^\T\sigma\eta\right\}=W(\eta)W(\xi)\exp\left\{-\frac\rmi 2\,\xi^\T\sigma\eta\right\}.
\end{equation}
More rigorously, the Weyl operators $W_1$ are defined as  projective unitary representations of the translation group $\Xi_1$ \cite{Hol01}, or as displacement operators acting on coherent vectors \cite{WisM10}.
Then, \eqref{WxiWxi'} represents the rigorous version of the CCR \cite{Hol01}.

\begin{remark}[Characteristic function of a state and Wigner function]\label{sec:Wigf}
As in the pure quantum case, the states $\hat \pi\in\Nscr_*$ are uniquely determined by their characteristic function $\chi_{\hat\pi}(\xi)$ \cite[Sec. 2.4]{DamWer23} or by their Wigner function $\Wcal_{\hat\pi}(z)$ \cite{WisM10}:
\begin{equation}\label{char+Wig}
\chi_{\hat\pi}(\xi)= \int_{\Rbb^s}\rmd x \,\Tr\left\{\hat\pi(x)W(\xi)(x)\right\}, \qquad \Wcal_{\hat \pi}(z)=\frac1{(2\pi)^{d}}\int_{\Xi}\rmd \xi\, \rme^{-\rmi z^\T \xi} \chi_{\hat\pi}(\xi).
\end{equation}

A function $\chi : \Xi \to \Cbb$ is the characteristic function of a state   \cite{DamWer23,BarW23}  if and only if
\\ (1) $\chi$ is continuous, \quad (2) $\chi(0)=1$,
\quad
(3) for every integer \ $N$ \ and every choice of \ $\xi_1, \ldots, \xi_N$, \ $\xi_j\in \Xi$, \  the $N\times N$-matrix with elements \  $\chi(\xi_k-\xi_l)\exp\left\{\frac \rmi 2 \,\xi_k^\T\sigma \xi_l\right\}$ \ is positive semi-definite,
\quad
(4) $\chi\in L^1(\Xi)$.
\end{remark}

\subsection{Quasi-free hybrid dynamical semigroup}\label{sec:defqfsemig}
\begin{definition}\label{def:2} A \emph{quasi-free hybrid dynamical semigroup} is a family of maps $\{\Tcal_t, \; t\geq 0\}$ such that Definition \ref{def:1} holds with $\Hscr=L^2(\Rbb^n)$ and
\begin{itemize}
\item[g.] (quasi-free property)
for all $\xi\in \Xi$
\begin{equation}\label{Tqf}
\Tcal_t[W(\xi)]=f_t(\xi)W(S_t\xi),
\end{equation}
where $S_t$
is a linear operator from $\Xi$ to $\Xi$, and $f_t$ is a continuous function from $ \Xi$ to $\Cbb$.
\end{itemize}
\end{definition}

The factor $f_t(\xi) $ is the \emph{noise function} and
$S_t$  gives the dynamics on the phase space. The main result of \cite{BarW23} concerns the explicit structure of these objects and the complete characterization of quasi-free hybrid dynamical semigroups.

\begin{theorem}\label{Mtheor} $\{\Tcal_t, \; t\geq 0\}$ satisfies Definitions \ref{def:1} and \ref{def:2}
if and only if $S_t$ and $f_t(\xi)$ have the following structure:
\begin{enumerate}
\item  $S_t=\rme^{Zt}, \ \forall t\geq 0$, \quad where $Z$  is a real $d\times d$-matrix;

\item  $f_t(\xi) = \exp\left(\int_0^t\rmd \tau \,\psi(S_\tau \xi)\right)$, \ where
\begin{equation}\label{psi}
\psi(\xi)=\rmi \alpha\cdot \xi -\frac 12 \,\xi\cdot A \xi+\int_{\Xi}\nu(\rmd \eta)\left(\rme^{\rmi {\eta}\cdot \xi}-1-\rmi \ind_{\left\{ \abs\eta<1\right\}}(\eta) {\eta}\cdot\xi\right) , \qquad \forall \xi\in \Xi=\Rbb^{d},
\end{equation}
$\alpha\in \Xi$, $A$ is a real symmetric $d\times d$-matrix with $A\geq 0$, $\ind_{\left\{ \abs\eta<1\right\}}$ is the indicator function of the sphere of radius 1, $\nu$ is  a $\sigma$-finite measure on $\Xi$, such that
\begin{equation}\label{nu}
\nu(\{0\})=0, \qquad \int_{\left\{ \abs\eta<1\right\}} \abs{\eta}^2\nu(\rmd \eta)<+\infty, \qquad \nu(\left\{ \abs\eta\geq 1\right\})<+\infty;
\end{equation}
\item
$A \pm \rmi B \geq 0$,\qquad $B:=\frac 1 2 \left(\sigma P_{1}Z-Z^\T P_{1}\sigma^\T\right)$.

\end{enumerate}

\end{theorem}
The super-script ${}^\T$ means matrix transposition and $P_{1}$ is the orthogonal projection on the quantum sector of the phase space:
\begin{equation}
P_{1}\Xi= \Xi_{1}, \qquad P_0=\openone- P_1, \quad P_{0}\Xi= \Xi_{0}.
\end{equation}

The structure of $\psi(\xi)$ is the classical \emph{L\'evy-Khintchine formula} and $\nu$ is known as
\emph{L\'evy measure}. The quantum features appear in the positivity condition (point 3): $\sigma$ comes from the CCR. The term with the indicator function in the integral has a compensating role and allows for measures $\nu$ with possible divergences in a  neighbourhood of zero. This compensating term can be written in different ways; the quantity $\psi$ can be left invariant by suitably changing $\alpha$.

\subsection{The master equation}
To better understand the dynamics of the hybrid system, it is useful to consider the master equation satisfied by the hybrid state. According to the discussion in Remark \ref{rem:Tstar}, the state at time $t$ is $\hat \pi_t={\Tcal_t}_*[\hat \pi_0]$; then, if $\Kcal$ is the generator of $\Tcal_t$ and $\Kcal_*$ its pre-adjoint, the state dynamics is given by the master equation
\begin{equation}
\frac{\rmd \ }{\rmd t}\,\hat \pi_t=\Kcal_*[\hat \pi_t].
\end{equation}
To express the structure of $\Kcal$ we firstly introduce some notation.

We write the matrices $A$ and $Z$ in the block form
\begin{equation}\label{blockAZ}
A=\begin{pmatrix} A^{11} & A^{10} \\ A^{01} & A^{00}\end{pmatrix}, \qquad Z=\begin{pmatrix} Z^{11} & Z^{10} \\ Z^{01} & Z^{00}\end{pmatrix},
\end{equation}
where $A^{11}$ is a real, non-negative $2n\times 2n$ matrix, $A^{00}$ is a real, non-negative $s\times s$ matrix, $A^{10}$ is a real $2n\times s$ matrix, $A^{01}= {A^{10}}^\T$,
$Z^{11}$ is a real $2n\times 2n$ matrix, $Z^{00}$ is a real $s\times s$ matrix, $Z^{10}$ is a real $2n\times s$ matrix, $Z^{01}$ is a real $s\times 2n$ matrix. We write also the vector $\alpha$ and the matrix $B$ in a similar block form:
\begin{equation}
\alpha = \begin{pmatrix} \beta \\ \alpha^0 \end{pmatrix}, \qquad B = \begin{pmatrix}B^{11}  & B^{10} \\ B^{01}& 0\end{pmatrix}=\frac 12 \begin{pmatrix}\sigma Z^{11} -{Z^{11}}^\T \sigma^\T & \sigma Z^{10} \\ -{Z^{10}}^\T\sigma^\T& 0\end{pmatrix}.
\end{equation}
Finally, we define
\begin{equation}\label{D<-Z}
D:=\frac 12 \left( Z^{11}\sigma +\sigma^\T{Z^{11}}^\T  \right),
\end{equation}
\begin{equation}
G:= \sigma^\T A^{11}\sigma +\frac \rmi 2 \left(\sigma^\T {Z^{11}}^\T - Z^{11}\sigma\right), \qquad C:= A^{00}, \qquad E:=\sigma^\T A^{10}-\frac \rmi 2\, Z^{10}.
\end{equation}
Then, the positivity condition 3. of Theorem \ref{Mtheor}, which is equivalent to $\left(\sigma^\T\otimes \openone\right) \left( A-\rmi B\right)\left(\sigma\otimes \openone\right) \geq 0$, can be written as
\begin{equation}\label{GEC>0}
\begin{pmatrix} G& E \\ E^\dagger & C\end{pmatrix}\geq 0.
\end{equation}
Let us stress that $Z^{01}$, $Z^{00}$, and  $D $ are not involved in the positivity condition.

By using the notation above and \eqref{R=QP}, it is easy to check that the generator $\Kcal$, given in \cite[Sec.\ 3]{BarW23}, can be rewritten in the following form, where the quantum-classical interaction terms are highlighted .

\begin{proposition}\label{prop:genVAR}
When $a$ is in the linear span of the Weyl operators, $f$ is bounded and twice differentiable, and $x_jf(x)$ ($j=1,\ldots,s$) is bounded, the generator $\Kcal$ of $\Tcal_t$ can be written as
\begin{equation}\label{Kcal}
\Kcal[a\otimes f](x)=f(x)\sum_{l=1}^2\Lcal_{\rm q}^l[a]+ a\sum_{l=1}^2\Kcal_{\rm cl}^l[ f](x) +\sum_{l=1}^4\Kcal_{\rm int}^l[a\otimes f](x),
\end{equation}
\begin{subequations}\label{Kcontrib}
\begin{equation}\label{Lq1}
\Lcal_{\rm q}^1[a]=\sum_{i,j=1}^{N}G_{ij}\left(R_iaR_j -\frac 12 \left\{R_iR_j,a\right\}\right)
+\rmi\left[ H_{\rm q},\, a\right] ,\qquad H_{\rm q}=\beta^\T  \sigma R +\frac 12 \,R^\T   D R,
\end{equation}
\begin{equation}\label{Lq2}
\Lcal_{\rm q}^2[a]=\int_{\Xi}\nu(\rmd \eta)\biggl\{W_1(\sigma\zeta)^\dagger a W_1(\sigma\zeta) - a
- \ind_{\left\{ \abs\eta<1\right\}}(\eta)  \rmi  \left[{\zeta}^\T \sigma R, \,a\right] \biggr\}, \qquad \eta=\begin{pmatrix}\zeta \\ y\end{pmatrix},
\end{equation}
\begin{equation}\label{K_cl1}
\Kcal_{\rm cl}^1[f](x)=  \sum_{j=1}^s \alpha^0_j\, \frac{\partial f(x)} {\partial x_j} + \sum_{i,j=1}^s x_i Z^{00}_{ij}\, \frac{\partial f(x)}{\partial x_j}+\frac 12\sum_{i,j=1}^{s}C_{ij}\,\frac{\partial^2 f(x)}{\partial x_i \partial x_j},
\end{equation}
\begin{equation}\label{K_cl2}
\Kcal_{\rm cl}^2[f](x)=  \int_{\Xi}\nu(\rmd \eta)\biggl\{f(x+y) - f(x)- \ind_{\left\{ \abs\eta<1\right\}}(\eta) \sum_{j=1}^s y_j\frac{\partial f(x) }{\partial x_j} \biggr\},  \qquad \eta=\begin{pmatrix}\zeta \\ y\end{pmatrix},
\end{equation}
\begin{equation}\label{K1+Hx}
\Kcal_{\rm int}^1[a\otimes f](x)=\rmi \left[H_x,\, a\right]f(x), \qquad H_x= x^\T  Z^{01}\sigma R,
\end{equation}
\begin{equation}\label{int2}
\Kcal_{\rm int}^2[a\otimes f](x)= - \sum_{i=1}^N \sum_{j=1}^{s}\left(\IM E_{ij}\right)\left\{R_i ,a \right\} \frac{\partial f(x)}{\partial x_j},\qquad \IM E_{ij}=-\frac 12 \, Z^{10}_{ij},
\end{equation}
\begin{equation}\label{int3}
\Kcal_{\rm int}^3[a\otimes f](x)= \rmi \sum_{i=1}^N \sum_{j=1}^{s} (\RE E_{ij})\left[R_i, a \right]\frac{\partial f(x)}{\partial x_j}, \qquad \RE E_{ij}= \left(\sigma^\T A^{10}\right)_{ij},
\end{equation}
\begin{equation}\label{int4}
\Kcal_{\rm int}^4[a\otimes f](x)=
\int_{\Xi}\nu(\rmd \eta)\left(f(x+y)-f(x)\right)\left(W_1(\sigma \zeta)^\dagger a W_1(\sigma \zeta)-a\right), \qquad \eta=\begin{pmatrix}\zeta \\ y\end{pmatrix}.
\end{equation}
\end{subequations}

\end{proposition}

The domain of the generator can be extended by linearity and weak$^*$-closure. Equation \eqref{Tqf} and Theorem \ref{Mtheor} give the explicit form of the action of $\Tcal_t$ on the Weyl operators; then, by linearity and weak$^*$-continuity, we obtained the action on the whole $W^*$-algebra $\Nscr$. So, the generator of the semigroup was not needed to determine the semigroup, but it is useful to better understand the dynamical behaviour and the physical interactions.

We have also separated the ``diffusive'' contributions $\Lcal_{\rm q}^1$, $\Kcal_{\rm cl}^1$, $\Kcal_{\rm int}^1$, $\Kcal_{\rm int}^2$ from the ``jump'' terms $\Lcal_{\rm q}^2$, $\Kcal_{\rm cl}^2$, $\Kcal_{\rm int}^3$, $\Kcal_{\rm int}^4$. As written at the end of Sec.\ \ref{sec:defqfsemig}, the compensating term in the jump part can be written in different ways, and this could change this separation.

\section{The structure of the quasi-free dynamics}\label{sec:qfdyn}
In the following we illustrate the role of the various terms introduced in Proposition \ref{prop:genVAR}.

\subsection{Reduced quantum dynamics} As said in Remark \ref{rem:state}, the reduced quantum state is
\[
\hat \rho_t= \int_{\Rbb^s} \rmd x\, \hat \pi_t(x).
\]
By using the notation \eqref{dualityform} for the duality form, we have
\[
\frac{\rmd \ }{\rmd t} \,\Tr\left\{\hat \rho_t a \right\}=\frac{\rmd \ }{\rmd t} \langle \hat \pi_t| a\otimes 1 \rangle =\langle \hat \pi_t| \Kcal[a\otimes 1] \rangle.
\]
So, the reduced quantum dynamics is obtained by taking $f(x)=1$ in the generator \eqref{prop:genVAR}; then, Eqs.\ \eqref{Kcontrib} give
\begin{equation}\label{qredK}
\Kcal[a\otimes 1](x)= \Lcal_{\rm q}^1[a] + \Lcal_{\rm q}^2[a]+\rmi[H_x,a].
\end{equation}

To have an autonomous reduced master equation, no $x$ dependence can appear in the reduced generator \eqref{qredK}; by \eqref{K1+Hx}, we must have $H_x=0$, i.e. $Z^{01}=0$. When $Z^{01}\neq 0$, the interaction term $\Kcal_{\rm int}^1$  can be seen as a random quantum Hamiltonian evolution, because the classical variables $x_j$, $j=1,\ldots,s$, appear in $H_x$.
We can say that $Z^{01}$ controls  the information flow from the classical system to the quantum one.

By construction, $\Lcal_{\rm q}^1+\Lcal_{\rm q}^2$ is the most general generator of a quasi-free quantum dynamical semigroup. Note that this unbounded generator has, formally, a structure of Lindblad type. This result was obtained in \cite{BarW23}, while generators with a non-vanishing ``jump'' part already appeared in the literature \cite{Hol96,VacH09}.

\subsection{Reduced classical dynamics}
Now, the classical reduced density and its dynamics are given by
\[
p_{t}(x)=\Tr\{\hat\pi_t(x)\}, \qquad \frac{\rmd \ }{\rmd t} \int_{\Rbb^s} \rmd x \,p_{t}(x) f(x)=\frac{\rmd \ }{\rmd t} \langle \hat \pi_t| \openone\otimes f \rangle =\langle \hat \pi_t| \Kcal[\openone\otimes f] \rangle.
\]
By Eqs.\ \eqref{Kcal}, \eqref{Kcontrib}, we get
\begin{equation}
\Kcal[\openone\otimes f](x)= \sum_{l=1}^2\Kcal_{\rm cl}^l[f](x)
+\Kcal_{\rm int}^2[\openone\otimes f](x), \qquad \Kcal_{\rm int}^2[\openone\otimes f](x)=\sum_{i=1}^N \sum_{j=1}^{s}  R_iZ^{10}_{ij}\,\frac{\partial f(x)}{\partial x_j}.
\end{equation}

Then, the reduced evolution equation of the classical component is autonomous only when $Z^{10}=0$. In this case we have
\begin{multline}
\frac{\rmd \ }{\rmd t}\, p_{t}(x)=-\sum_{j=1}^s \alpha^0_j\, \frac{\partial p_{t}(x)} {\partial x_j} - \sum_{i,j=1}^s  Z^{00}_{ij}\, \frac{\partial x_ip_{t}(x)}{\partial x_j}+\frac 12\sum_{i,j=1}^{s}C_{ij}\,\frac{\partial^2 p_{t}(x)}{\partial x_i \partial x_j}
\\ {}+\int_{\Xi}\nu(\rmd \eta)\biggl\{p_{t}(x-y) - p_{t}(x)+ \ind_{\left\{ \abs\eta<1\right\}}(\eta) \sum_{j=1}^s y_j\frac{\partial p_{t}(x) }{\partial x_j} \biggr\},  \qquad \eta=\begin{pmatrix}\zeta \\ y\end{pmatrix}.
\end{multline}
This is a version of the Kolmogorov-Fokker-Planck equation \cite[Secs.\ 3.5.2, 3.5.3]{App09}, giving rise to  semigroups of transition probabilities of time-homogeneous Markov processes \cite{Sato99,App09}. For $C=0$ and $\nu=0$, one can also obtain the Liouville equation for a system with quadratic Hamiltonian \cite[Sec.\ 5.1.2]{BarW23}.

\subsection{Dissipation and information exchange}
In principle a classical system can be observed without disturbing its dynamics. Due to the clas\-sical-quantum interaction, by observing the classical component one can gain information on the quantum subsystem without changing the dynamics of the total system. However, this flow of information is possible only when some dissipation is present in the dynamics. The usual statement that ``measurements perturb a quantum system'' becomes something like ``some irreversibility must be present in the quantum dynamics to extract information from a quantum system''. In the modern formulation of quantum measurement theory the notions of observables and state reduction have been generalized by the notions of \emph{positive operator valued measures} (also called \emph{resolution of identity}) and \emph{instruments} (or \emph{operation} valued measures) \cite{Hol01,WisM10,BarG09}, which indeed we have already constructed in Sec.\ \ref{sec:instr}.

In the following subsections \ref{sec:nodissq}, \ref{sec:nodissc} we shall show how some irreversibility and noise source are needed to have a non-trivial quantum-classical dynamics. However, we are working only in the quasi-free case, while this point was raised also in other approaches, see for instance \cite{Opp+23,Dio23}.

\subsubsection{No dissipation in the quantum system}\label{sec:nodissq}
We consider now the case of no dissipation in the quantum subsystem, in the sense that the reduced quantum dynamics is of purely Hamiltonian type. So, we take $G=0$ in \eqref{Lq1} to have $\Lcal_{\rm q}^1$ of purely Hamiltonian type; then, we need $\Lcal_{\rm q}^2=0$ and we take the measure $\nu$ concentrated on $\Xi_0$: $\int_\Xi\nu(\rmd \zeta,\rmd y)g(\zeta,y)=\int_{\Xi_0} \mu(\rmd y) g(0,y)$.
By the positivity condition \eqref{GEC>0} we get also $ E=0$; then, we have
\[
\Lcal_{\rm q}^2=0, \qquad \Kcal_{\rm int}^l=0 \ \ \text{for} \ l=2,3,4.
\]
The term $\Kcal_{\rm cl}^1$ remains unchanged and $\Kcal_{\rm cl}^2$ becomes
\begin{equation*}
\Kcal_{\rm cl}^2[f](x)=  \int_{\Xi_0}\mu(\rmd y)\biggl\{f(x+y) - f(x)- \ind_{\left\{ \abs y<1\right\}}(y) \sum_{j=1}^s y_j\frac{\partial f(x) }{\partial x_j} \biggr\}.
\end{equation*}
Finally, the total generator \eqref{Kcal} reduces to
\begin{equation*}
\Kcal[a\otimes f](x)=\rmi\left[ H_{\rm q}+H_x,\, a\right]f(x)+ a\sum_{l=1}^2\Kcal_{\rm cl}^l[ f](x) .
\end{equation*}
Only a single interaction term survives, a Hamiltonian term which gives a force exerted by the classical system on the quantum one.

Without dissipation in the quantum system, there is no possibility for the classical system to extract information from the quantum component. Some ``smooth state reduction'' is needed to extract information from a quantum system.

\subsubsection{No dissipation in the classical system}\label{sec:nodissc}
A similar situation happens when we ask for no dissipation in the classical component. Let us assume $C=0$ and $\Kcal_{\rm cl}^2=0$, which gives again $E=0$; moreover, the measure $\nu$ turns out to be concentrated on $\Xi_1$:
\[
\int_\Xi\nu(\rmd \zeta,\rmd y)g(\zeta,y)=\int_{\Xi_1} \tilde \mu(\rmd \zeta) g(\zeta,0).
\]
Then, we have again $\Kcal_{\rm int}^l=0$ for $l=2,3,4$. Also in this situation no information can flow from the quantum system to the classical one.

When some quantum information is extracted, its intrinsic probabilistic character introduces a certain degree of uncertainty in the classical signal.

\subsection{The interaction terms}

The interaction $\Kcal_{\rm int}^1$ \eqref{K1+Hx} involves the random Hamiltonian $H_x$; it is the only interaction term which appears in the reduced quantum dynamics \eqref{qredK}. This term represents a  force exerted on the quantum system by the classical one.

On the other side, $\Kcal_{\rm int}^2$ \eqref{int2} is the only  interaction  surviving in the reduced classical dynamics and it represents some action of the quantum system on the classical one.
The matrix $\IM E_{ij}$, appearing in this interaction, is involved in the positivity condition \eqref{GEC>0}; we can say that this interaction term injects some quantum uncertainty into the classical output.

The interaction terms $\Kcal_{\rm int}^3$ \eqref{int3} and $\Kcal_{\rm int}^4$ \eqref{int4} have a peculiar structure, as they vanish either when the reduced classical dynamics is considered ($a=\openone$), either when the reduced quantum dynamics is considered ($f(x)=1$). Their effects are visible only in the total joint dynamics and modify the quantum-classical correlations.

In the general theory of measurements in continuous time and in previous examples of hybrid dynamics, some stochastic evolution equations for quantum states have been introduced \cite{WisM10,BarH95,Dio23,Opp+23,BarB91,BarG13,BarG09}. They give the \emph{conditional state}, which means the state to be attributed to the quantum system at time $t$, when the trajectory of the classical system is known up to $t$. This construction is also known as stochastic unraveling; in this framework, the physical meaning of the interaction terms (mainly of the last two) should become more transparent.

\section{Possible developments} \label{sec:compare}

As already discussed, the notion of hybrid dynamics of a quantum-classical system is connected to measurements in continuous time. In the Markovian case quantum continuous measurements have been formalized in \cite{BarHL93,BarP96} by the notion of \emph{semigroup of probability operators}. Such semigroups are defined by the properties of Definition \ref{def:1}, to which the translation invariance of the classical component is added, in order to represent a signal without an intrinsic dynamics.
The translations in the classical component are defined by: $\forall f\in L^\infty(\Rbb^s)$,  $\Rcal_z[f](x)=f(x+z)$ (almost everywhere); we shall identify $\Rcal_z$ and $\id\otimes \Rcal_z$. Then, a \emph{semigroup of probability operators} is a hybrid dynamical semigroup, as given in Definition \ref{def:1}, to which the invariance restriction is added:
\begin{itemize}
\item[h.] $\Rcal_z\circ \Tcal_t=\Tcal_t\circ \Rcal_z$, \qquad $\forall z\in\Rbb^s$, \quad $\forall t\geq 0$.
\end{itemize}
In \cite{BarP96} the generator of the most general semigroup of probability operators has been found under a further continuity restriction, which implies that only bounded operators on the quantum component are involved in this generator. The construction of the generator is again based on the L\'evy-Khintchine formula. Essentially, in the final expression the quantum position and momentum operators $R_j$ \eqref{R=QP} are substituted by generic bounded operators on $\Hscr$. Moreover, the jump part has again  an integral structure very similar to the one of the quasi-free case, but now not only unitary operators can appear in this integral. Since the semigroups found in \cite{BarP96} are in a sub-class of the hybrid semigroups, one could try to modify them to construct more general non-quasi free hybrid semigroups.

To better understand how to proceed, it is useful to see what happens in the quasi-free case (Def.\ \ref{def:2}) by adding the restriction h. By using \eqref{Tqf} we have
\[
(\Rcal_z\circ\Tcal_t)[W(\xi)](x)=f_t(\xi)W_1(P_1S_t\xi)W_0(P_0S_t\xi)(x+z),
\]
\[
(\Tcal_t\circ\Rcal_z)[W(\xi)](x)=f_t(\xi)W_1(P_1S_t\xi)(x)W_0(P_0S_t\xi)(x)\rme^{\rmi P_0\xi\cdot z};
\]
by applying the restriction h.\, this gives $P_0S_t\xi=P_0\xi$, $\forall\xi\in \Xi=\Xi_1\oplus \Xi_0$. So, by differentiating with respect to $t$ and using the block structure \eqref{blockAZ}, we get
\[
Z^{00}=0, \qquad Z^{01}=0.
\]
The same result can be obtained by asking the commutation of classical translations with the various terms \eqref{Kcontrib} of the generator. The restriction  $Z^{00}=0$ affects only the classical dynamics
$\Kcal_{\rm cl}^1$ \eqref{K_cl1} and gives the vanishing of the deterministic part of the classical motion. The restriction $Z^{01}=0$ is equivalent to $\Kcal_{\rm int}^1=0$, which means the vanishing of the force exerted by the classical system on the quantum one. No other change appears in the generator.
It can be checked that the final form of $\Kcal$ is, formally, in the class of the generators obtained in \cite{BarP96}, but with position and momentum operators instead of bounded operators on $\Hscr$. We see that the effect of adding the translation invariance is to suppress the only two terms were the quantity $x$ appears: $H_x$ and $\sum_{i,j=1}^s x_i Z^{00}_{ij}\, \frac{\partial f(x)}{\partial x_j}$. So, a possibility to get most general hybrid semigroup is to allow for a $x$-dependence in the various terms of the generator of \cite{BarP96}.

A second way to get other classes of hybrid dynamical semigroups is to use stochastic differential equations in Hilbert spaces, as done in \cite{Dio23,Opp+23,BarH95}. While the construction of \cite{Dio23,Opp+23} was done with the explicit aim of constructing examples of quantum-classical dynamics, the approach of \cite{BarH95} was developed having in mind the study of a class of stochastic differential equations and the generalization of continuous measurements to non-Markovian cases. In any case, in the Markovian case the expression \cite[(4.39)]{BarH95} is found, which indeed modifies  the generator of \cite{BarP96} by introducing many $x$-dependencies both in the diffusive part and in the jump part. Apart the restriction that only bounded operators on $\Hscr$ are allowed, this expression gives rise to a very general master equation for a quantum-classical dynamics.

Finally, a fruitful approach to quantum measurements in continuous time has been through the use of quantum stochastic calculus \cite{Bar06,BarG13,Hol01}; this should open new possibilities also to construct quantum-classical dynamical theories.


\begin{thebibliography}{99}
\bibitem{DGS00} L. Di\'osi, N. Gisin,  W.T. Strunz, \textsl{Quantum approach to coupling classical and quantum dynamics}, Phys. Rev. A \textbf{61} (2000) 022108. \href{https://doi.org/10.1103/PhysRevA.61.022108}{DOI: 10.1103/PhysRevA.61.022108}
\bibitem{Dio14} L. Di\'osi, \textsl{Hybrid quantum-classical master equations}, Phys. Scr. \textbf{T163} (2014) 014004. \href{https://doi.org/10.1088/0031-8949/2014/T163/014004}{DOI: 10.1088/0031-8949/2014/T163/014004}
\bibitem{Dio23} L. Di\'osi, \textsl{Hybrid completely positive Markovian quantum-classical dynamics},  Phys. Rev. A \textbf{107} (2023)  062206. \href{https://doi.org/10.1103/PhysRevA.107.062206}{DOI: 10.1103/PhysRevA.107.062206}
\bibitem{ManRT23} G. Manfredi, A. Rittaud, C. Tronci, \textsl{Hybrid quantum-classical dynamics of pure-dephasing systems}, J. Phys. A: Math. Theor. \textbf{56} (2023) 154002. \href{https://doi.org/10.1088/1751-8121/acc21e}{DOI: 10.1088/1751-8121/acc21e}

\bibitem{Opp+23} J. Oppenheim, C. Sparaciari, B. Šoda, Z. Weller-Davies, \textit{Objective trajectories in hybrid classical-quantum dynamics},
Quantum \textbf{7} (2023)  891. \href{https://doi.org/10.22331/q-2023-01-03-891}{DOI: 10.22331/q-2023-01-03-891}

\bibitem{OppWD22} J. Oppenheim, Z. Weller-Davies, \textit{The constraints of post-quantum classical gravity}, J. High Energ. Phys. 2022 J. High Energ. Phys. 2022, 80 (2022). \href{https://doi.org/10.1007/JHEP02(2022)080 (2022)}{DOI: 10.1007/JHEP02(2022)080}
\bibitem{ProsB23}    G.M. Prosperi, M. Baldicchi, \textit{Interpretation of Quantum Theory and Cosmology}, arXiv:2304.07095 (2023). \href{https://doi.org/10.48550/arXiv.2304.07095}{DOI: 10.48550/arXiv.2304.07095}

\bibitem{Sergi+23} A. Sergi, D. Lamberto, A. Migliore, A. Messina, \textsl{Quantum–classical hybrid systems and Ehrenfest's Theorem}, Entropy \textbf{25} (2023) 602. \href{https://doi.org/10.3390/e25040602}{DOI: 10.3390/e25040602}
\bibitem{Pomar+23} J. L. Alonso, C. Bouthelier-Madre, J. Clemente-Gallardo, D. Mart\'inez-Crespo, J. Pomar, \textsl{Effective nonlinear Ehrenfest hybrid quantum-classical dynamics}, Eur. Phys. J. Plus \textbf{138} (2023) 649. \href{https://doi.org/10.1140/epjp/s13360-023-04266-w}{DOI: 10.1140/epjp/s13360-023-04266-w}
\bibitem{BarL05} A.~Barchielli, G.~Lupieri,
    \textsl{Instruments  and channels in quantum information theory}, Optika i Spektroskopiya
    \textbf{99} (2005) 443--450; Optics and Spectroscopy \textbf{99} (2005)
    425--432. \href{https://link.springer.com/article/10.1134/1.2055938}{DOI: 10.1134/1.2055938}
\bibitem{DamWer23} L. Dammeier, R.F. Werner, \textsl{Quantum-classical hybrid systems and their quasifree transformations}, Quantum \textbf{7} (2023) 1068. \href{https://doi.org/10.22331/q-2023-07-26-1068}{DOI: 10.22331/q-2023-07-26-1068}
\bibitem{BarW23} A. Barchielli, R. Werner, \textsl{Hybrid quantum-classical systems: Quasi-free Markovian dynamics}, arXiv:2307.02611 (2023). \href{https://doi.org/10.48550/arXiv.2307.02611}{DOI: 10.48550/arXiv.2307.02611}

\bibitem{BarB91} A.~Barchielli, V.~P.~Belavkin, \textsl{Measurements
    continuous in time and   a posteriori states in quantum mechanics},
    J.\ Phys.\ A: Math.\ Gen. {\bf24} (1991) 1495--1514. \href{http://iopscience.iop.org/0305-4470/24/7/022}{DOI:  10.1088/0305-4470/24/7/022}
\bibitem{BarHL93} A.\ Barchielli, A.\ S.\ Holevo, G.\ Lupieri, \textsl{An  analogue of Hunt's representation theorem in quantum probability}, J.  Theor. Probab. {\bf 6} (1993) 231--265. \href{https://doi.org/10.1007/BF01047573}{DOI: 10.1007/BF01047573}
\bibitem{BarP96} A.~Barchielli, A.~M.~Paganoni, \textsl{A
    note on a formula of L\'evy--Khinchin type in quantum probability}, Nagoya Math. J. {\bf141} (1996) 29--43.
\bibitem{BarG09} A.~Barchielli, M.~Gregoratti,
    \href{http://www.springer.com/physics/quantum+physics/book/978-3-642-01297-6}{\textit{Quantum Trajectories and Measurements in Continuous Time: The Diffusive Case}}, Lect.\ Notes Phys.\ \textbf{782} (Springer, Berlin \& Heidelberg,  2009).
\bibitem{Bar06} A. Barchielli, \textsl{Continual Measurements in Quantum Mechanics and  Quantum Stochastic Calculus}.
    In S. Attal, A. Joye, C.-A. Pillet (eds.),
    \textit{Open  Quantum Systems III}, Lect.\ Notes Math.\ \textbf{1882}
    (Springer, Berlin, 2006), pp. 207--291. \href{https://link.springer.com/chapter/10.1007/3-540-33967-1_5}{DOI: 10.1007/b128453}
\bibitem{BarG13} A.\ Barchielli, M.\ Gregoratti, \textsl{Quantum continuous
    measurements: The stochastic Schr\"odinger equations and the spectrum
    of the output},  Quantum Measurements and Quantum Metrology, \textbf{1}
    (2013) 34--56.  \href{http://dx.doi.org/10.2478/qmetro-2013-0005}{DOI:
    10.2478/qmetro-2013-0005}

\bibitem{Van78} P. Vanheuverzwijn, \textsl{Generators for quasi-free completely positive semi-groups}, \href{http://www.numdam.org/item/AIHPA_1978__29_1_123_0/}{Ann. I. H. P. A \textbf{29} (1978) 123--138}.
\bibitem{DVV79} B. Demoen, P. Vanheuverzwijn, A. Verbeure, \textsl{Completely positive quasi-free maps of the CCR-algebra}, Rep.\ Math.\ Phys.\ \textbf{15} (1979) 27--39. \href{https://doi.org/10.1016/0034-4877(79)90049-1}{DOI: 10.1016/0034-4877(79)90049-1}
\bibitem{Hell10} M. Hellmich, \textsl{Quasi-free semigroups on the CCR algebra}, Rep. Math. Phys. \textbf{66} (2010) 277--298. \href{https://doi.org/10.1016/S0034-4877(10)80031-X}{DOI: 10.1016/S0034-4877(10)80031-X}
 \href{https://doi.org/10.1017/S0027763000005511}{DOI: 10.1017/S0027763000005511}
\bibitem{Sato99} K. Sato, \textit{L\'evy processes and infinitely divisible distributions} (Cambridge University Press, Cambridge, 1999).
\bibitem{Hol01} A.S. Holevo, \textit{Statistical Structure of Quantum Theory}, Lecture Notes in Physics m 67 (Springer, Berlin, 2001).
\bibitem{WisM10} H.M. Wiseman and G.J. Milburn, \textit{Quantum Measurement and Control} (Cambridge University Press, Cambridge, 2010).
\bibitem{BarH95} A.~Barchielli, A.~S.~Holevo, \textsl{Constructing
    quantum measurement processes via classical stochastic calculus},  Stoch. Proc. Appl. {\bf 58} (1995) 293--317. \href{http://dx.doi.org/10.1016/0304-4149(95)00011-U}{DOI: 10.1016/0304-4149(95)00011-U}

\bibitem{Hol96} A.S. Holevo, \textsl{Covariant quantum Markovian evolutions}, J. Math. Phys. \textbf{37} (1996) 1812--1832. \href{https://pubs.aip.org/aip/jmp/article-abstract/37/4/1812/307394/Covariant-quantum-Markovian-evolutions?redirectedFrom=fulltext}{DOI: https://doi.org/10.1063/1.531481}
\bibitem{VacH09} B. Vacchini, K. Hornberger, \textsl{Quantum linear Boltzmann equation}, Phys. Rep. \textbf{478} (2009) 71--120. \href{https://www.sciencedirect.com/science/article/abs/pii/S0370157309001422}{DOI: 10.1016/j.physrep.2009.06.001}
\bibitem{App09} D. Applebaum, \textit{L\'evy Processes and Stochastic Calculus},
Second Edition (Cambridge University Press, 2009).

\end{thebibliography}
\end{document}